 \documentclass[prb,twocolumn,showpacs,preprintnumbers,amsmath,amssymb, superscriptaddress]{revtex4}

\usepackage{graphicx}
\usepackage{dcolumn}
\usepackage{bm}
\usepackage{color}

\begin{document}

\newcommand{\siox}{SiO$_2$}
\newcommand{\silicate}{Si$_2$O$_3$}
\newcommand{\sicoxint}{\mbox{SiC/SiO$_2$}}
\newcommand{\rootthree}{($\sqrt{3}$$\times$$\sqrt{3}$)R30$^{\circ}$}
\newcommand{\rootthreehl}{($\sqrt{\mathbf{3}}$$\times$$\sqrt{\mathbf{3}}$)R30$^{\circ}$}
\newcommand{\sixroot}{(6$\sqrt{3}$$\times$6$\sqrt{3}$)R30$^{\circ}$}
\newcommand{\sixroothl}{(6$\sqrt{\mathbf{3}}$$\times$6$\sqrt{\mathbf{3}}$)R30$^{\circ}$}
\newcommand{\three}{\mbox{(3${\times}$3)}}
\newcommand{\four}{\mbox{(4${\times}$4)}}
\newcommand{\five}{\mbox{(5${\times}$5)}}
\newcommand{\six}{\mbox{(6${\times}$6)}}
\newcommand{\two}{\mbox{(2${\times}$2)}}
\newcommand{\twobyc}{\mbox{(2$\times$2)$_\mathrm{C}$}}
\newcommand{\twobysi}{\mbox{(2$\times$2)$_\mathrm{Si}$}}
\newcommand{\seven}{($\sqrt{7}$$\times$$\sqrt{7}$)R19.1$^{\circ}$}
\newcommand{\fourseven}{(4$\sqrt{7}$$\times$4$\sqrt{7}$)R19.1$^{\circ}$}
\newcommand{\one}{\mbox{(1${\times}$1)}}
\newcommand{\sicbar}{SiC(000$\bar{1}$)}
\newcommand{\sicbarhl}{SiC(000$\bar{\mathbf{1}}$)}
\newcommand{\bardir}{(000$\bar{1}$)}
\newcommand{\grad}{\mbox{$^{\circ}$}}
\newcommand{\cgrad}{\,$^{\circ}$C}
\newcommand{\projecta}{(11$\bar{2}$0)}
\newcommand{\projectb}{(10$\bar{1}$0}
\newcommand{\projectc}{(01$\bar{1}$0}
\newcommand{\third}{$\frac{1}{3}$}
\newcommand{\thirdspot}{($\frac{1}{3}$,$\frac{1}{3}$)}
\newcommand{\oversix}{$\frac{1}{6}$}
\newcommand{\kpoint}{$\bar{\textrm{K}}$-point}
\newcommand{\kpar}{\underline{k}$_{\parallel}$}


\title{Band structure engineering of epitaxial graphene on SiC by molecular doping}

\author{C. Coletti}
\email{c.coletti@fkf.mpg.de}
\affiliation{Max-Planck-Institut f\"{u}r Festk\"{o}rperforschung, Heisenbergstr. 1, D-70569 Stuttgart}
\author{C. Riedl}
\affiliation{Max-Planck-Institut f\"{u}r Festk\"{o}rperforschung, Heisenbergstr. 1, D-70569 Stuttgart}
\author{D.S. Lee}
\affiliation{Max-Planck-Institut f\"{u}r Festk\"{o}rperforschung, Heisenbergstr. 1, D-70569 Stuttgart}
\author{B. Krauss}
\affiliation{Max-Planck-Institut f\"{u}r Festk\"{o}rperforschung, Heisenbergstr. 1, D-70569 Stuttgart}
\author{L. Patthey}
\affiliation{Paul Scherrer Institut, CH-5232 Villigen-PSI, Switzerland}
\author{K. von Klitzing}
\affiliation{Max-Planck-Institut f\"{u}r Festk\"{o}rperforschung, Heisenbergstr. 1, D-70569 Stuttgart}
\author{J.H. Smet}
\affiliation{Max-Planck-Institut f\"{u}r Festk\"{o}rperforschung, Heisenbergstr. 1, D-70569 Stuttgart}
\author{U. Starke}%
\email{u.starke@fkf.mpg.de}
\homepage{http://www.fkf.mpg.de/ga}
\affiliation{Max-Planck-Institut f\"{u}r Festk\"{o}rperforschung, Heisenbergstr. 1, D-70569 Stuttgart}

\date{\today}

\begin{abstract}
Epitaxial graphene on SiC(0001) suffers from strong intrinsic n-type doping.
We demonstrate that the excess negative charge can be fully compensated by non-covalently functionalizing graphene with the strong electron acceptor tetrafluorotetracyanoquinodimethane (F4-TCNQ). Charge neutrality can be reached in monolayer graphene as shown in electron dispersion spectra from angular resolved photoemission spectroscopy (ARPES). In bilayer graphene the band gap that originates from the SiC/graphene interface dipole increases with increasing F4-TCNQ deposition and, as a consequence of the molecular doping, the Fermi level is shifted into the band gap. The reduction of the charge carrier density upon molecular deposition is quantified using electronic Fermi surfaces and Raman spectroscopy. The structural and electronic characteristics of the graphene/F4-TCNQ charge transfer complex are investigated by X-ray photoelectron spectroscopy (XPS) and ultraviolet photoelectron spectroscopy (UPS). The doping effect on graphene is preserved in air and is temperature resistant up to 200 {\cgrad}. Furthermore, graphene non-covalent functionalization with F4-TCNQ can be implemented not only via evaporation in ultra-high vacuum but also by wet chemistry.
\end{abstract}

\pacs{Valid PACS appear here}

\maketitle

\section{Introduction}
\label{intro} The electronic properties of graphene, such as large
room temperature mobilities, comparable conductivities for electrons
and holes and the ability for charge
carrier operation via the field effect, make it an excellent
candidate for carbon based
nanoelectronics~\cite{Novoselov2004, Novoselov2005, Zhang2005}.
However, the limited size of graphene flakes from conventional
micromechanical cleaving~\cite{Novoselov2004} requires individual
selection and handling which makes device fabrication cumbersome. In
contrast, epitaxial graphene grown on silicon carbide (SiC) offers
realistic prospects for large scale graphene
samples~\cite{Berger2006, Riedl2007, Emtsev2009}. Unfortunately,
as-grown epitaxial graphene is electron doped as a result of the
graphene/SiC interface properties~\cite{Ohta2006, Ohta2007, Zhou2007, Riedl2008, Bostwick2007}. This doping translates into a
displacement of the Fermi energy, E$_\textrm{F}$, away from the
Dirac point energy E$_\textrm{D}$ where the $\pi$-bands
cross, so that the ambipolar properties of graphene
cannot be exploited. Several approaches can be used to remove or compensate this excess charge. One that has recently been introduced is the structural decoupling of the graphene layers from the substrate using hydrogen
intercalation~\cite{Riedl2009}. Also, chemical gating techniques are
very promising to tune the carrier concentration as
demonstrated recently in low temperature
experiments on graphene flakes~\cite{Wehling2008, Lohmann2009}. Analogously, a
possibility to compensate the n-doping in epitaxial graphene is to extract the surplus negative carriers, i.e. --- in the language of semiconductors --- to
accomplish a method of hole injection.

Similar to the case of carbon
nanotubes~\cite{Kong2000, Collins2000}, injection of holes in
graphene can be achieved via surface adsorption of gas molecules
such as O$_{2}$ or the paramagnetic NO$_{2}$~\cite{Zhou2008,
Schedin2007}. In contrast, NH$_{3}$ and alkali metals such
as potassium are known to act as electron donors
in carbon based materials~\cite{Kong2000, Schedin2007, Derycke2002,
Ohta2006}. However, the high reactivity of NO$_{2}$, NH$_{3}$ and of
alkali atoms makes those materials ill-suited as practical dopants.
This is illustrated by the need of cryogenic temperatures and ultra
high vacuum conditions to stably adsorb NO$_{2}$ and potassium on
graphene surfaces~\cite{Zhou2008, Ohta2006}. An approach that
promises to control the carrier type and concentration in graphene
in a simple and reliable way is that of surface transfer doping via
organic molecules~\cite{Chen2009}. A variety of aromatic and
non-aromatic molecules and even organic free radicals can be used to
control graphene doping~\cite{Su2009, Bekyarova2009, Chen2007,
Lu2009, Choi2010}. Many of these molecules possess good thermal
stability, have limited volatility after adsorption and can be easily applied via wet chemistry. An effective p-type dopant is the strong electron
acceptor tetrafluoro-tetracyanoquinodimethane (F4-TCNQ). It has a
very high electron affinity (i.e., E$_\textrm{ea}$ = 5.24 eV) and
has been used successfully as a state of the art p-type dopant in
organic light emitting diodes~\cite{Chen2009, Blockwitz1998,
Zhou2001, Gao2008}, carbon nanotubes~\cite{Kazaoui2003,
Takenobu2005, Nosho2007} and on other materials~\cite{Gao2001,
Gao2002}. Recently, the existence of a p-doping
effect of F4-TCNQ on graphene has been suggested
theoretically~\cite{Pinto2009} and experimentally~\cite{Chen2007}.

\begin{figure*}
\begin{center}
\includegraphics[width=163mm]{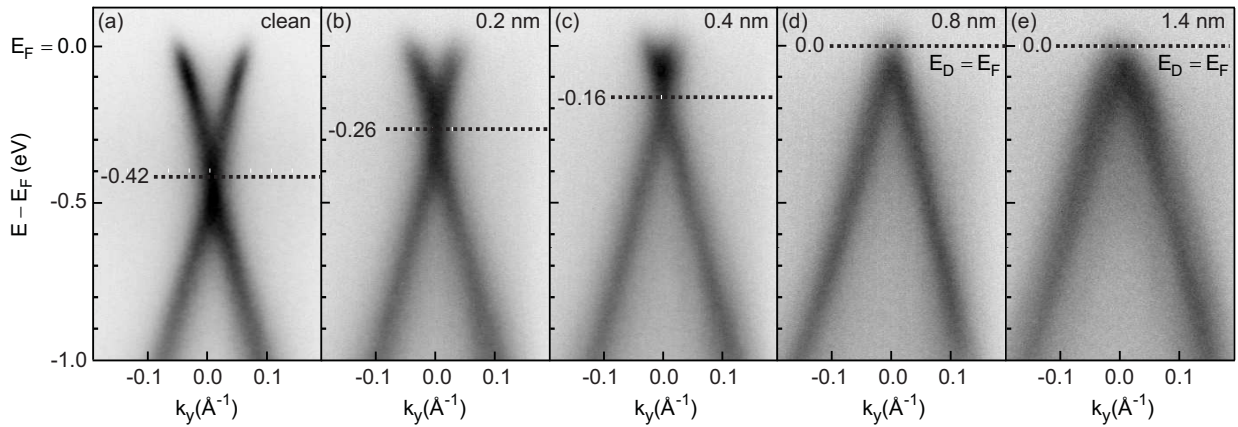}
\end{center}
\caption{Dispersion of the $\pi$-bands measured with UV excited
ARPES around the $\bar{K}$ point of the graphene
Brillouin zone for (a) an as-grown graphene monolayer on SiC(0001)
and (b-e) for the same sample covered with an increasing
amount of F4-TCNQ molecules. The momentum scans are taken
perpendicular to the $\bar{\Gamma}\bar{K}$-direction in reciprocal
space. The Fermi level E$_\textrm{F}$ shifts progressively towards
the Dirac point (E$_\textrm{D}$, dotted black line) with increasing nominal
thickness of the deposited F4-TCNQ film. Charge neutrality
(E$_\textrm{F}$ = E$_\textrm{D}$) is reached for a molecular coverage of 0.8 nm (d). When depositing additional
molecules the Fermi level does not shift any further (e).} \label{arpes1}
\end{figure*}

In the present paper we give direct evidence that the excess negative charge in epitaxial monolayer graphene can be fully compensated by functionalizing its surface with F4-TCNQ. Electron dispersion spectra and Fermi surface maps measured via angle resolved photoemission spectroscopy (ARPES) qualitatively and quantitatively evaluate the reduction in charge carrier density and show that charge neutral graphene can be ultimately obtained. X-ray photoelectron spectroscopy (XPS) and ultraviolet photoelectron spectroscopy (UPS) elucidate the structural and electronic characteristics of the graphene/F4-TCNQ charge transfer complex. Raman spectroscopy of the G phonon peak corroborates the doping reversal and shows that the carrier concentration can be trimmed by laser induced desorption of molecules. Moreover, we investigate the effects of F4-TCNQ on the band structure of bilayer graphene. By presenting a band gap ~\cite{Ohta2006, Ohta2007, Zhou2007, Riedl2008}, bilayer graphene is particularly attractive for the implementation of electronic devices such as field effect transistors provided that the intrinsic doping can be compensated. Here we demonstrate that F4-TCNQ not only renders bilayer graphene semiconducting thanks to the full compensation of the excess negative charged carriers but also increases the band gap size to more than double of its initial value. We show that the molecular layer is stable when exposed to air. The doping effect is preserved up to 200 {\cgrad} and is totally reversible by annealing the sample at higher temperatures. The molecular coverage can be precisely controlled when using a
molecular evaporator but the dopants can also be applied by wet
chemistry, i.e. in a technologically convenient way.

\section{Experimental}
\label{exp}

\begin{figure}
\begin{center}
\includegraphics[width=77.361mm]{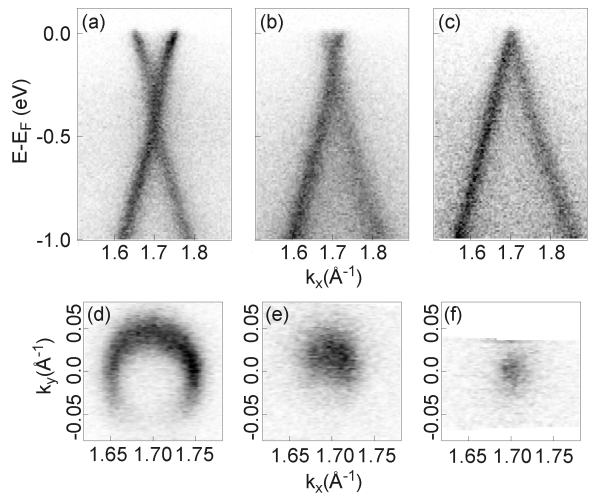}
\end{center}
\caption{Dispersion of the $\pi$-band around the $\bar{K}$
point of the graphene Brillouin zone measured by ARPES with
synchrotron light in scans oriented parallel to
the $\bar{\Gamma}\bar{K}$-direction for (a) a
pristine epitaxial graphene monolayer, (b) an
intermediate F4-TCNQ coverage and (c) the F4-TCNQ coverage leading to
charge neutrality. Panels (d) through (f) show the corresponding constant energy maps at E$_\textrm{F}$. From these Fermi surface maps we extract a charge carrier concentration of 7.3 $\pm$ 0.2 $\cdot$10$^{12}$ cm$^{-2}$ for the pristine graphene,
9 $\pm$ 2 $\cdot$10$^{11}$ cm$^{-2}$ for the intermediate coverage and 1.5 $\pm$ 2 $\cdot$10$^{11}$ cm$^{-2}$ for full coverage. All the spectra shown were acquired with circular polarized light with a photon energy of 30 eV and at a sample temperature of 80 K.} \label{arpesSLS}
\end{figure}

Epitaxial graphene was grown in UHV by thermal Si
sublimation~\cite{Riedl2007} on hydrogen
etched~\cite{Soubatch2005,Frewin2009}, atomically-flat 6H-SiC(0001)
crystals. The samples were characterized with low energy electron
diffraction (LEED) and angular resolved photoemission spectroscopy
(ARPES). Subsequently, F4-TCNQ molecules
(7,7,8,8-Tetracyano-2,3,5,6-tetrafluoroquinodimethane, Sigma
Aldrich, 97\% purity) were deposited on the graphene/SiC substrates
by thermal evaporation from a resistively-heated crucible. For
comparison also the non-fluorinated version of F4-TCNQ, i.e.
tetracyanoquinodimethane (TCNQ) was deposited
(7,7,8,8-tetracyanoquinodimethane, Sigma Aldrich, 98\% purity).
In house ARPES measurements were carried out at
room temperature (RT) using monochromatic He II
radiation ($h\nu$ = 40.8 eV) from a UV discharge source with a
display analyzer oriented for momentum scans perpendicular to the
$\bar{\Gamma}\bar{K}$-direction of the graphene Brillouin zone.
The Fermi surface data were extracted from ARPES
experiments using synchrotron radiation from the Swiss Light Source (SLS) of the Paul Scherrer Institut (PSI), Switzerland, at the Surface and Interface Spectroscopy beamline (SIS). The endstation allows, using a display analyzer and a sample manipulator with three rotational degrees of freedom, for fast high-resolution two-dimensional electronic dispersion measurements. XPS measurements were performed
using photons from a non-monochromatic Mg
K$_{\alpha}$ source ($h\nu$ = 1253.6 eV). The stability of the
molecular layers under UV and X-ray irradiation was verified by
exposing 3 hours and well over 13 hours, respectively. The thickness
of the deposited molecular layers was estimated from XPS spectra
calibrated through a comparison to spectra for a well characterized
surface phase of TCNQ on Cu(100) measured under identical
conditions~\cite{Tseng2009}. Different deposition rates ranging from
0.07 to 0.5 \AA/min and sample temperatures between -140 and 25
{\cgrad} were tested for the sample preparation. No influence on the
doping results was found when the same amount of molecules was
deposited. Work function measurements and the analysis of molecular orbitals
were performed via normal emission UPS using
monochromatic He I radiation ($h\nu$ = 21.21 eV)
from our UV source. During the work function measurements a bias of
-30V was applied to the sample in order to distinguish between the
analyzer and the sample cut-off and to more efficiently collect the
inelastically scattered low kinetic energy electrons into the
analyzer. Raman spectra were measured under ambient conditions using
an Argon ion laser with a wavelength of 488 nm at a power level of
12 mW and a laser spot size of $\approx$ 1 $\mu$m in diameter. In
order to apply the molecular layer on graphene via wet chemistry
F4-TCNQ was dissolved in either chloroform or
dimethyl sulfoxide (DMSO) until saturation. Before ARPES
characterization the sample was left immersed in the solution for 12
hours.

\section{F4-TCNQ on monolayer graphene}
\label{results1}

\begin{figure*}
\begin{center}
\includegraphics[width=154.584mm]{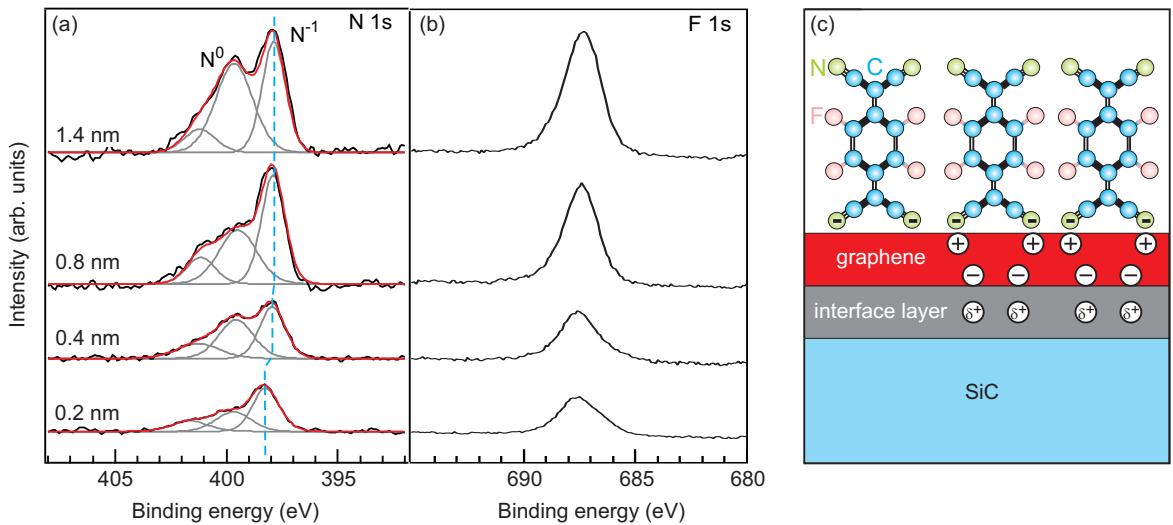}
\end{center}
\caption{(a,b) XPS spectra of the N 1s (a) and F 1s (b) core level
emission regions from submonolayer (bottom spectrum) to multilayer
(top spectrum) amounts of F4-TCNQ deposited on a monolayer of graphene which has been grown epitaxially on SiC(0001). Three different components are fitted into
the N 1s region and are assigned to N$^{-1}$ and
N$^{0}$ species and to a shake-up process. The blue dashed line
indicates the exact energy position of the N$^{-1}$ component as it
shifts with molecular layer thickness. (c) Schematic structure of a
F4-TCNQ layer deposited on top of a graphene layer grown on SiC. The
charges induced in the graphene layer due to the interface dipole
and the molecular charge transfer are indicated.} \label{xps}
\end{figure*}

The doping level of the graphene layers can be precisely monitored
with ARPES measurements of the $\pi$-band dispersion around the
$\bar{K}$-point of the graphene Brillouin zone as
previously established~\cite{Ohta2006, Ohta2007, Zhou2007,
Riedl2008, Bostwick2007}. As shown in Fig. \ref{arpes1}(a) for an
as-grown monolayer of graphene on SiC(0001) the Fermi level
E$_\textrm{F}$ is located about 0.42 eV above the
Dirac point E$_\textrm{D}$. This corresponds to the
well established charge carrier concentration value of n $\approx$ 1
x 10$^{13}$ cm$^{-2}$ for as grown graphene. For increasing amounts of deposited F4-TCNQ E$_\textrm{F}$ moves back towards E$_\textrm{D}$ as illustrated in
Fig. \ref{arpes1}(b)-(d). Meanwhile the bands remain sharp, which
indicates that the integrity of the graphene layer is preserved.
Evidently, deposition of F4-TCNQ activates electron transfer from
graphene towards the molecule thus neutralizing the excess doping
induced by the substrate. As the figure shows the electron
concentration in the graphene layer can be tuned precisely by
varying the amount of deposited molecules. When we deposit a 0.8 nm
thick layer of molecules, charge neutrality is reached, i.e.
E$_\textrm{F}$ = E$_\textrm{D}$. For a nominal thickness of the
molecular film above 0.8 nm no additional shift of the Fermi energy
is observed as seen in Fig. \ref{arpes1}(e), which indicates that
the charge transfer saturates.

For a detailed quantitative determination of the
carrier concentrations, high-resolution ARPES data acquired using
synchrotron radiation were analyzed. Fig. ~\ref{arpesSLS} compares
the $\pi$-band dispersion (a-c) and constant energy maps (d-f) at
E$_\textrm{F}$ for a clean graphene monolayer (a,d), an intermediate
F4-TCNQ coverage (b,e) and charge transfer saturation at full
coverage (c,f). The charge carrier concentration
can be derived precisely from the size of the Fermi surface pockets as $n = (k_F - k_{\bar{\textrm{K}}})^2/\pi$, where
$k_{\bar{\textrm{K}}}$ denotes the wave vector at the boundary of
the graphene Brillouin zone. The Fermi surface pocket radius is extracted by using Lorentzian fits of the maxima of the momentum distribution curves of the electronic dispersion spectra in panels (a-c). The corresponding
carrier concentrations are 7.3$\cdot$10$^{12}$ cm$^{-2}$, 9$\cdot$10$^{11}$
cm$^{-2}$ and 1.5$\cdot$10$^{11}$ cm$^{-2}$ for the clean graphene
monolayer, the intermediate and the higher coverage, respectively. The error bar for the reported carrier concentrations is $\pm$ 2$\cdot$10$^{11}$
cm$^{-2}$ and was determined from the variance of the Lorentzian fits.

\begin{figure}
\begin{center}
\includegraphics[width=85mm]{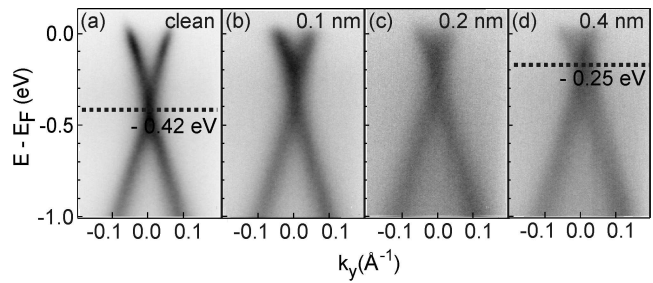}
\end{center}
\caption{Dispersion of the $\pi$-bands measured with ARPES through
the $\bar{K}$-point of the graphene Brillouin zone for (a) a pristine graphene monolayer grown on SiC(0001) and (b-d) for increasing amounts of TCNQ deposited on graphene. The Fermi level (E$_\textrm{F}$) shifts progressively
towards the Dirac point (E$_\textrm{D}$, black dotted line) for
increasing molecular coverage up to a value of E$_\textrm{D}$ -
E$_\textrm{F}$ = - 0.25 eV.} \label{arpes2}
\end{figure}

\section{Characterization of the charge transfer complex}
\label{results2}

The location of the charge transfer process within the F4-TCNQ
molecule can be elucidated by core level analysis using XPS. N 1s
and F 1s core level emission spectra for different amounts of
deposited F4-TCNQ are displayed in Fig. \ref{xps}. For the N 1s
spectra of panel (a) a line shape analysis reveals two main
components centered at binding energies (BE) of 398.3 and 399.6 eV. This indicates that different N species exist in the
deposited molecular film. In agreement with the
literature~\cite{Wells1991, Chen2007} the peak at 398.3 eV is
assigned to the anionic species N$^{-1}$ while the 399.6 eV
component is attributed to the neutral N$^{0}$ species. The
additional broad component at 401.7 eV likely originates from
shake-up processes in view of its energy location
and the relative intensity (approximately 20\%) as compared to the main
peak~\cite{Lindquist1998}. The F 1s spectra in Fig. \ref{xps}(b) are
in contrast dominated by a single component. Only at low coverages a
weak asymmetry develops. The appearance of the N$^{-1}$ anion
species indicates that the electron transfer takes place through the
C$\equiv$N groups of the molecules while the fluorine atoms are
largely inactive. A similar mechanism with electronically active
cyano groups has been found for F4-TCNQ on other
surfaces\cite{Lindquist1998, Romaner2007, Qi2007}. However, in the
present case not all C$\equiv$N groups are involved in the charge
transfer process. While for low molecular coverages the N$^{-1}$ species
dominate (71\%), for coverages from 0.4 nm to 0.8
nm about 45\% of the C$\equiv$N groups are uncharged (N$^{0}$) as
determined from the peak areas (0th momentum) of the fitted
components. This indicates that when the films are densely packed,
most of the molecules are standing upright as sketched in Fig.
\ref{xps}(c) (apparently, in dilute layers not all molecules are
arranged perpendicular to the surface). We note, that this result
is only valid for the initial molecular layer and is different
than what was recently proposed for multilayers (5 nm) of
F4-TCNQ~\cite{Chen2009}. The energy position of the different core
level peaks shifts with increasing molecular coverage as indicated
by the blue dashed line in Fig. \ref{xps}(a). For 0.8 nm nominal film
thickness this shift is exactly the same as the shift of the
$\pi$-bands with respect to the Fermi energy E$_\textrm{F}$ (i.e.
0.4 eV for saturation) in agreement with our working hypothesis of a
strong electronic coupling between the F4-TCNQ molecule and the
graphene surface. At coverages larger than 0.8 nm, the shift
of both the N$^{-1}$ peak and the band structure saturates. Only the
N$^{0}$ peak continues to grow indicating the formation of a charge
neutral second layer of molecules. The saturation effect at 0.8 nm
nominal film thickness also supports the model of a dense layer of upright
standing molecules since the size of an F4-TCNQ molecule along its
axis is indeed about 0.8 nm. 

A comparison of the experimental band
shifts when using the non-fluorinated version of the F4-TCNQ
molecule, i.e. tetracyanoquinodimethane (TCNQ), shows that the
charge transfer is greatly enhanced when the F species are present,
even though they are not directly involved in the charge transfer
process. With TCNQ, which has a much smaller electron affinity than
F4-TCNQ (i.e., 2.8 eV for TCNQ compared to 5.24 eV for F4-TCNQ), the
Fermi energy remains at least 0.25 eV above the Dirac point (see
Fig. \ref{arpes2}). The maximum shift of the band structure measured
upon TCNQ deposition is obtained for a molecular coverage of 0.4 nm (see Fig. \ref{arpes2}(d)) and no additional shift is
observed for higher amounts of deposited molecules. 

\begin{figure}
\begin{center}
\includegraphics[width=85mm]{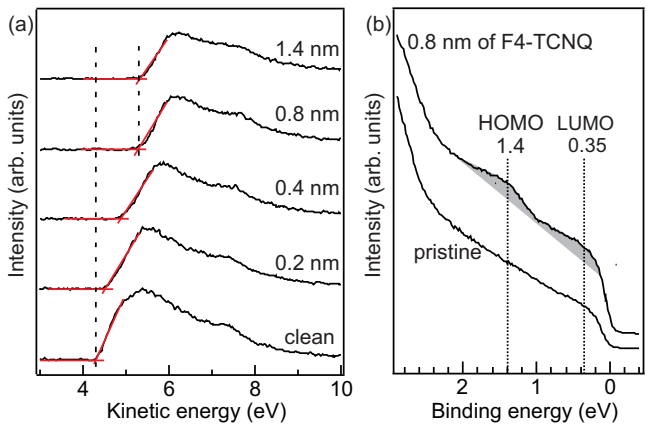}
\end{center}
\caption{(a) Secondary electron cutoff region measured
by normal emission UPS (h$\nu$ = 21.21 eV) for
increasing nominal thickness of a F4-TCNQ film deposited on
epitaxial monolayer graphene on SiC(0001) in order to estimate the
work function change ($\Delta\Phi$). (b) Near E$_\textrm{F}$ UPS
spectra for clean graphene (bottom) and graphene with 0.8 nm of deposited F4-TCNQ
(top). The shaded areas highlight the emerging features following
F4-TCNQ deposition and are attributed to the HOMO (at 1.4 eV) and to
the LUMO which has partially shifted below E$_\textrm{F}$ (at 0.35
eV).} \label{ups}
\end{figure}

Additional
evidence for the formation of charge transfer complexes in the case
of F4-TCNQ is obtained from the work function measurements shown in
Fig. \ref{ups}(a). The kinetic energies are plotted after correction
for the applied bias and the analyzer work function, so that the
sample work function is directly obtained from the intersection
between the base line of the spectrum and a linear fit to the tail
of the sample secondary electron cut-off. The work function ($\Phi$)
gradually increases from 4.28 eV for as-grown epitaxial graphene to
a maximum value of 5.29 eV for 0.8 nm of F4-TCNQ on top of graphene
and saturates for larger molecular coverages. The measured shift
($\Delta\Phi$ $\approx$ 1 eV) contains both the band bending at the
graphene surface (0.4 eV) and an additional contribution from the
interface dipole generated by the charge transfer (i.e. $\approx$
0.6 eV). 

An analysis of the
position of the highest occupied (HOMO) and lowest unoccupied (LUMO)
molecular orbitals of F4-TCNQ with respect to the Fermi level using normal emission UPS corroborates further
that the molecule gets charged. The
low BE portion of the UPS spectra of a graphene
sample with a 0.8 nm molecular coverage exhibits two additional shoulders,
which are not observed for pristine epitaxial graphene. They are located
at 1.4 eV and 0.35 eV (see Fig. \ref{ups}(b)). In agreement with the
literature~\cite{Chen2007,Gao2002, Koch2000}, the higher BE peak is
attributed to the HOMO and the lowest BE peak to the (now partially
populated) LUMO of the molecule. Even though the HOMO of the
pristine molecule is typically found at higher BE
values~\cite{Koch2005} and the LUMO is expected for negative BE
values, filling of the former LUMO of F4-TCNQ with one electron
generates a negative polaron~\cite{Koch2000}. Hence, the LUMO
is stabilized, i.e. the binding energy of the newly occupied
state is increased. In contrast, the former HOMO is destabilized (lower BE).

\section{Raman spectroscopy analysis}

\begin{figure}
\begin{center}
\includegraphics[width=86mm]{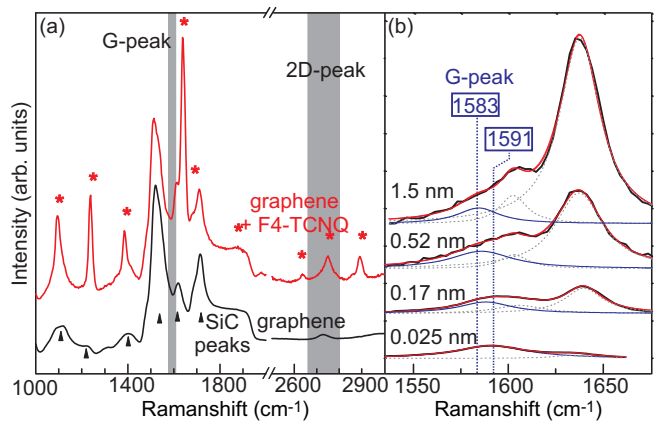}
\end{center}
\caption{(a) Raman spectrum of pristine (bottom black trace) and
F4-TCNQ-modified (top red trace) monolayer
graphene epitaxially grown on SiC(0001). Molecular peaks are marked with stars and the peaks related to SiC with arrows. The G- and
2D-peak regions of graphene are shaded. (b) Differential Raman spectra for different coverages of the F4-TCNQ molecular film ranging from 1.5 nm to 0.025 nm (see main text). The grey dashed lines are
Lorentzians to fit the molecular peaks. The blue solid line is the
extracted graphene contribution to the Raman spectrum (G-peak).} 
\label{raman1}
\end{figure}

\begin{figure*}
\begin{center}
\includegraphics[width=164mm]{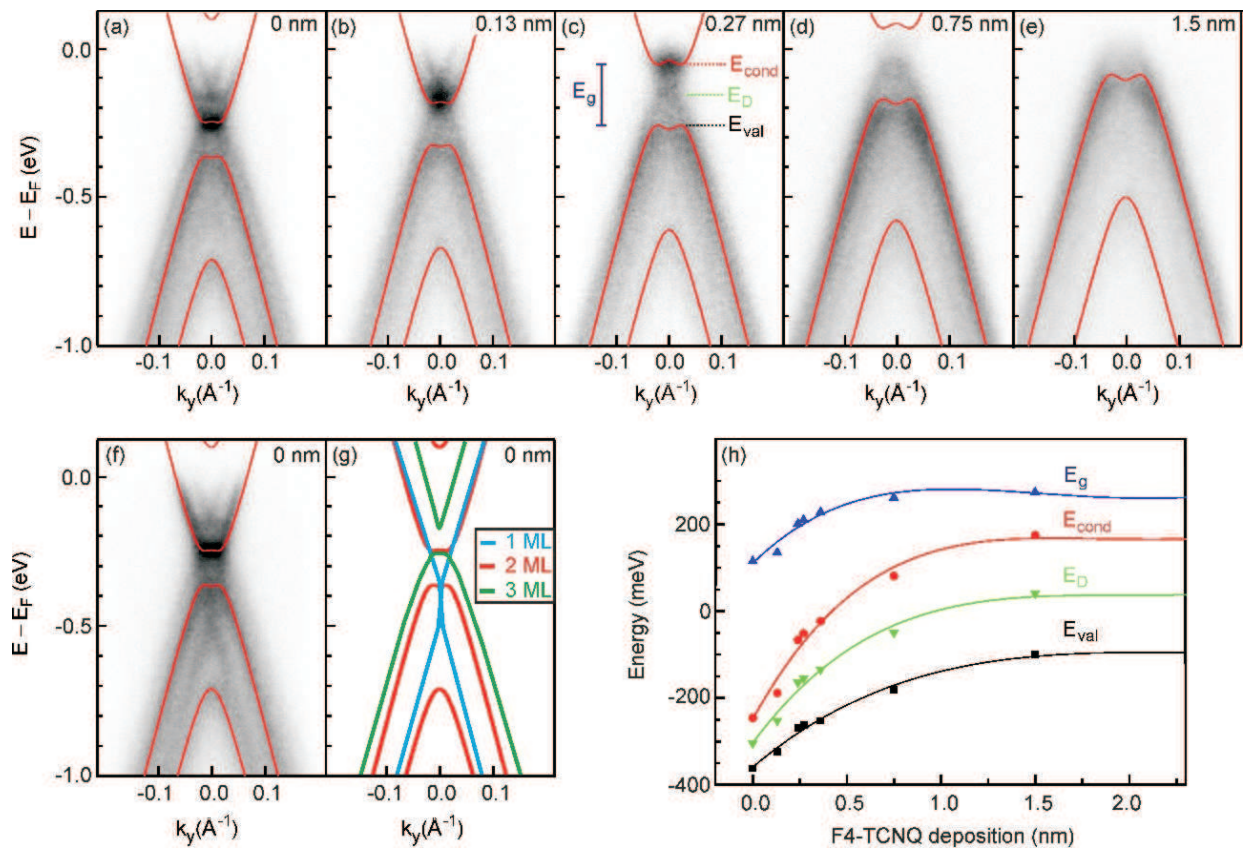}
\end{center}
\caption{ARPES band structure plots measured perpendicular to the
$\bar{\Gamma}\bar{K}$-direction for an epitaxially grown graphene bilayer on
SiC(0001) (a) without F4-TCNQ coverage and
(b-e) with increasing amounts of F4-TCNQ. Bands calculated
within a tight binding model are superimposed to the experimental
data. (f) ARPES data showing the band structure of an epitaxial
graphene bilayer prepared at a lower annealing temperature.
Contributions from monolayer domains are evident. (g) Schematic band
structure of mono-, bi-, and trilayer epitaxial graphene. (h)
Evolution of the energy gap E$_\textrm{g}$, the gap midpoint or
Dirac point E$_\textrm{D}$, the minimum of the lowest conduction
band E$_\textrm{cond}$ and the maximum of the uppermost valence band
E$_\textrm{val}$ as a function of molecular coverage. The evolution of the energies for higher molecular coverages (up to 5 nm, not shown) confirms charge transfer saturation. The definition
of the energies is included in panel (c).} 
\label{arpes3}
\end{figure*}

The influence of the F4-TCNQ coverage on the vibrational and
electronic properties of the graphene layer was also studied under
ambient conditions with Raman spectroscopy. Figure \ref{raman1}(a)
compares Raman spectra for an as-grown epitaxial monolayer of
graphene (bottom trace) and for a sample that has been covered with
a 1.5 nm thick F4-TCNQ layer (top trace). Peaks related to the SiC
substrate are marked by arrows. The 2D-peak of graphene is
highlighted with grey shading. So is the G-peak. The latter is
barely visible due to overwhelming contributions of the SiC
substrate in this wavelength range~\cite{Lee2008}. The Raman
spectrum for graphene covered with F4-TCNQ reveals numerous
additional features that are marked by stars. By illuminating a
sample that is covered with F4-TCNQ molecules with the
Argon laser light it is possible to gradually remove the
deposited molecules through evaporation. Features associated with
the SiC substrate and graphene hardly change, while the peaks
attributed to the F4-TCNQ molecules decrease in amplitude. Laser
heating can therefore be used to trim the molecule coverage and
hence tune the charge carrier concentration in graphene. In a
confocal arrangement it is therefore possible to spatially modulate the
doping level. The charge carrier concentration can be extracted from
a detailed inspection of the G-peak. In order to eliminate the large
contributions of the SiC substrate, it is instrumental to analyze
differential spectra obtained by subtracting the Raman data of the
clean hydrogen-etched SiC substrate from the spectrum of the F4-TCNQ-modified
graphene layer on top of SiC\cite{Lee2008,Roehrl2008}. The evolution
of the G-peak upon successive laser illumination, i.e. for
successively reduced amounts of F4-TCNQ, is illustrated in Fig.
\ref{raman1}(b). Only the spectral region from 1530 to 1700
cm$^{-1}$ centered around the G-peak is shown. The spectra can be decomposed into three peaks. Two 
molecular peaks at $\approx$ 1602 and $\approx$ 1637 cm$^{-1}$
decrease with the laser exposure. The molecular coverage before laser exposure was calibrated with XPS (top curve in panel (b)). The other {molecular coverages} marked in Fig. \ref{raman1}(b) are calculated from the relative intensity of the molecular peaks. The intensity of the remaining
peak, which we attribute to the G phonons of graphene, is
approximately constant and not influenced by laser exposure. The
peak position shifts however from $\approx$ 1583 to $\approx$ 1591
cm$^{-1}$. In graphene the carrier density enters the electron
phonon coupling and causes phonon stiffening when the carrier
density increases. The G-peak position of the F4-TCNQ saturated
sample (1583.3 $\pm$ 0.9 cm$^{-1}$) is nearly the same as for charge
neutral graphene flakes~\cite{Yan2007, Das2008}. This is consistent
with the ARPES data. As the molecules are successively removed, the
G-peak blue-shifts and finally reaches 1591 cm$^{-1}$, the value
for clean monolayer graphene on SiC exposed to air~\cite{Lee2008}. This G-peak position corresponds to a charge carrier concentration of $\approx$ 5 x
10$^{12}$ cm$^{-2}$\,\,~\cite{Yan2007, Das2008} or a band gap shift of
E$_\textrm{F}$ - E$_\textrm{D}$ $\approx$ 0.3 eV. We note, that this
value is less than measured by ARPES (E$_\textrm{F}$ -
E$_\textrm{D}$ = 0.42 eV) due to the additional doping when the sample is exposed to air as reported previously~\cite{Lee2008}. 

\section{F4-TCNQ on bilayer graphene}

For bilayers the band shift caused by the intrinsic
n-doping of epitaxial graphene on SiC is slightly lower than for
epitaxial monolayers, namely about 0.3 eV. In addition, the
electric dipole present at the graphene/SiC interface imposes an
electrostatic asymmetry between the layers which causes a band gap
to open by roughly 0.1 eV~\cite{Ohta2006, Ohta2007,
Zhou2007, Riedl2008} as seen from the ARPES data in Fig.
\ref{arpes3}(a). In the figure bands obtained from
tight-binding calculations are superimposed to the dispersion plot.
This facilitates an analytical evaluation of the Dirac energy
position and the size of the band gap. The calculations are based on
a symmetric bilayer Hamiltonian as described  by McCann and
Fal'ko~\cite{McCann2006}. We note that, due to the inevitable
inhomogeneity of UHV-prepared graphene samples and the beam spot
size, the ARPES data contain contributions of film areas with
different thickness. This can be seen by a comparison with data from
a sample prepared at a slightly lower temperature in Fig.
\ref{arpes3}(f). Here, the contribution from monolayer patches is
notably stronger and obstructs a clear view on the bilayer bands.
The sketch in panel (g) identifies the band contributions stemming
from different graphene thicknesses. In the sample used for panel
(a) the bilayer bands are well isolated, although trilayer
contributions are clearly present. Similar to the monolayer case,
F4-TCNQ deposition onto this sample causes a progressive shift of
the bilayer bands, i.e. a reduction of the intrinsic n-type doping.
This is illustrated in the measured and calculated
dispersion plots in Fig. \ref{arpes3}(b)-(e). Concurrent with the
drop of E$_\textrm{F}$-E$_\textrm{D}$, the size of the band gap
increases as seen from the bands fitted with the tight binding
simulations. The band fitting retrieves the energy
at the bottom of the lowest conduction band E$_\textrm{cond}$ and at
the top of the uppermost valence band E$_\textrm{val}$. From these
values the energy gap E$_\textrm{g}$ and the mid gap or Dirac energy
E$_\textrm{D}$ are derived. The corresponding energies are marked in
panel (c). The evolution of the characteristic energies of these
fitted bands with the amount of deposited molecules is plotted in
Fig. \ref{arpes3}(h). The band gap E$_\textrm{g}$ increases from 116
meV for a clean as-grown bilayer to 275 meV when a 1.5 nm thick
layer of F4-TCNQ molecules has been deposited. We
verified that no further charge transfer occurs for higher amounts
of deposited molecules. The Fermi energy moves into the
band gap for a molecular layer thickness of 0.4 nm. Hence the bilayer is turned from a conducting system into a truly semiconducting layer. The increase of
the band gap indicates that the molecular deposition increases the
on-site Coulomb potential difference between both layers. From the
tight binding calculations we get an increase in the on-site Coulomb
interaction from 0.12 eV for a clean bilayer to 0.29 eV for a
bilayer with a molecular coverage of 1.5 nm~\cite{tb}. This
increase can be attributed to an increased electrostatic field due
to the additional dipole developing at the graphene/F4-TCNQ
interface.

\begin{figure}
\begin{center}
\includegraphics[width=63.5mm]{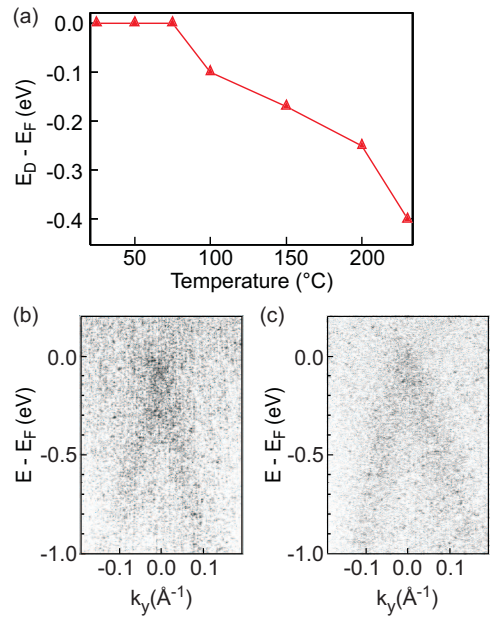}
\end{center}
\caption{(a) Shift of the Fermi level E$_\textrm{F}$ with respect to
the Dirac energy E$_\textrm{D}$ as a function of
temperature during annealing of F4-TCNQ covered epitaxial monolayer
graphene on SiC(0001). The shift was determined from ARPES data
recorded after each annealing step in UHV of a graphene sample
with an initial molecular coverage of 1.5 nm.
(b) Incomplete shift of the $\pi$-band dispersion
after F4-TCNQ wet chemical application in chloroform. (c) $\pi$-band
dispersion after F4-TCNQ wet chemical application in DMSO. Charge
neutrality is achieved, as indicated by E$_\textrm{F}$ =
E$_\textrm{D}$.} \label{wet}
\end{figure}

\section{Thermal stability and chemical application
of the molecules}

An important aspect of the F4-TCNQ/graphene system is the robustness
of its preparation: the Raman experiments after transport through
ambient environment already demonstrated that the charge transfer
complex is stable in air. On a monolayer sample covered with a
multilayer of F4-TCNQ molecules the band structure was measured with
ARPES before and after several hours of air exposure. This
experiment revealed no change in the band structure. XPS
measurements also confirmed the inert nature of the graphene
substrate. The experiment with laser light exposure suggests that
the F4-TCNQ layer is sensitive to temperature. The volatility of
F4-TCNQ was probed in UHV by stepwise annealing a sample with a
molecular coverage of 1.5 nm. The sample was annealed repeatedly
for 1 min at successively higher temperatures between 25 {\cgrad} to
230 {\cgrad} in steps of about 25 degrees. After each annealing step
the shift of the Fermi level E$_\textrm{F}$ with respect to the
Dirac energy E$_\textrm{D}$ was determined from ARPES spectra
recorded at room temperature. As the annealing temperature increased
the difference between the Dirac energy and the Fermi energy
increased back to the value of a pristine graphene layer. This
increase is considered direct evidence for molecular desorption from
the graphene surface. As is evident from Fig. \ref{wet}(a),
desorption of the molecules is initiated at temperatures around 75
{\cgrad} and completed at 230 {\cgrad}. Since thermal desorption is
amplified by UHV conditions we anticipate that even higher
temperatures are needed under atmospheric pressure to remove the
entire molecular layer. Finally, we demonstrate that the F4-TCNQ
layer can also be applied by immersing the sample in a chemical
F4-TCNQ solution. Two solvents were tested to apply the molecular
layer on graphene via wet chemistry: chloroform and dimethyl
sulfoxide (DMSO). ARPES spectra taken immediately after
introduction into UHV show a considerable background due to
contamination by residual chemicals from the solution as displayed
in Fig. \ref{wet}(b) and (c). Nevertheless, the shift
of the band structure is clearly visible, and in the case of F4-TCNQ wet chemical application in DMSO (panel (c))
charge neutrality (i.e., E$_\textrm{F}$ = E$_\textrm{D}$) is
achieved.

\section{Conclusion}

\label{concl} In conclusion, we have demonstrated that the band
structure of epitaxial graphene on SiC(0001) can be
precisely tailored by functionalizing the graphene surface with
F4-TCNQ molecules. Charge neutrality can be achieved for mono- and
bilayer graphene. A charge transfer complex is
formed by the graphene film and the F4-TCNQ molecular overlayer. The electrons are removed from the graphene layer
via the cyano groups of the molecule.  Since the
molecules remain stable under ambient conditions, at elevated temperatures and can be applied via wet chemistry this doping method is attractive as its incorporation into existing technological processes appears feasible. In bilayer graphene, the hole doping allows the Fermi
level to shift into the energy band gap and the additional dipole developing
at the interface with the F4-TCNQ overlayer causes the
band gap magnitude to increase to more than double of its original
value. Thus, the electronic structure of the graphene bilayer can be precisely tuned by varying the molecular coverage.

\begin{acknowledgments}
The authors thank C.L. Frewin, C. Locke and S.E. Saddow of the University
of South Florida for hydrogen etching of the SiC substrates. C.C.
acknowledges the Alexander von Humboldt research fellowship for
financial support. The research leading to these results has received funding from the European Community's Seventh Framework Programme (FP7/2007-2013) under grant agreement no. 226716.
\end{acknowledgments}

\end{document}